# CitNetExplorer: A new software tool for analyzing and visualizing citation networks

Nees Jan van Eck and Ludo Waltman

Centre for Science and Technology Studies, Leiden University, The Netherlands
{ecknjpvan, waltmanlr}@cwts.leidenuniv.nl

We present CitNetExplorer, a new software tool for analyzing and visualizing citation networks of scientific publications. CitNetExplorer can for instance be used to study the development of a research field, to delineate the literature on a research topic, and to support literature reviewing. We first introduce the main concepts that need to be understood when working with CitNetExplorer. We then demonstrate CitNetExplorer by using the tool to analyze the scientometric literature and the literature on community detection in networks. Finally, we discuss some technical details on the construction, visualization, and analysis of citation networks in CitNetExplorer.

## 1. Introduction

A lot of work has been done on the analysis and visualization of many different types of bibliometric networks (Börner, 2010; Van Eck & Waltman, in press). Analyses for instance have focused on networks of co-authoring researchers or networks of keywords co-occurring in publications. However, the most frequently studied types of bibliometric networks are based on citation relations. Examples include networks of co-citation relations or bibliographic coupling relations between journals, researchers, or individual publications.

The analysis and visualization of direct citation networks has received relatively limited attention in the scientometric literature. Interest has focused much more on co-citation and bibliographic coupling networks than on direct citation networks. Many techniques have been developed for analyzing and visualizing co-citation and bibliographic coupling networks. Also, various software tools are available to support the study of these networks (Cobo, López-Herrera, Herrera-Viedma, & Herrera, 2011;

Van Eck & Waltman, in press), among which our own VOSviewer tool (see www.vosviewer.com; Van Eck & Waltman, 2010).

Important work on the analysis and visualization of direct citation networks has been done by Eugene Garfield and colleagues (Garfield, Pudovkin, & Istomin, 2003; see also Garfield, 2004, 2009). Garfield emphasizes the value of direct citation networks for studying the history and development of research fields. He refers to this as algorithmic historiography. Garfield has developed a software tool called HistCite that can be used to construct and visualize direct citation networks based on data downloaded from Thomson Reuters' Web of Science database. Nowadays, HistCite is made freely available for personal use by Thomson Reuters (see www.histcite.com).

In this paper, we introduce CitNetExplorer, a new software tool that we have developed for analyzing and visualizing direct citation networks. CitNetExplorer, which is an abbreviation of 'citation network explorer', builds on Garfield's work on algorithmic historiography. Compared with HistCite, CitNetExplorer can handle much larger citation networks, possibly including millions of publications and citation relations. Moreover, CitNetExplorer offers sophisticated functionality for drilling down into a citation network, for instance allowing users to start at the level of a full network consisting of several millions of publications and to then gradually drill down into this network until a small subnetwork has been reached including no more than, say, 100 publications, all dealing with a specific topic of interest. CitNetExplorer borrows various ideas from our VOSviewer tool. This applies in particular to certain features related to visualization (e.g., smart labeling) and user interaction (e.g., zooming and panning). CitNetExplorer can be downloaded from www.citnetexplorer.nl. The tool has been developed in Java and therefore should run on any system that offers Java support.

For what types of purposes can CitNetExplorer be used? Below, we give some examples. The first two examples can be seen as a kind of macro-level applications of CitNetExplorer, while the last two examples apply more at the micro level.

*Studying the development of a research field over time*. This is what is referred to as algorithmic historiography by Garfield. The idea is that by showing the most important publications in a field, ordered by the year in which they appeared, and the citation relations between these publications, one obtains a picture of the development of a field over time. Notice that more traditional analyses based on co-citation and bibliographic coupling networks yield a static picture of a research field and therefore do not make clear how a field has evolved over time.

*Delineating research areas*. Suppose one wants to identify all publications on a certain research topic or in a certain research area. One may attempt to identify publications using keywords or based on the journal in which they have appeared, but this will usually yield an incomplete result. Publications that do not contain the right keywords will be missed. The same applies to publications that have not appeared in the right journal. Starting from a core set of relevant publications, CitNetExplorer can be used to identify publications based on citation relations. One may for instance select all publications that have at least a certain minimum number of citation relations with publications in the core set.

*Studying the publication oeuvre of a researcher*. At a more micro level, CitNetExplorer can be used to study the publication oeuvre of an individual researcher. By showing citation relations between the publications of a researcher, it becomes clear how publications build on each other and possibly how the oeuvre of a researcher consists of multiple more or less independent parts, each dealing with a different research topic. In addition, based on citation relations, CitNetExplorer can show the earlier literature on which a researcher builds in his work and, the other way around, the more recent literature that in some way has been influenced by the work of a researcher.

*Literature reviewing*. Literature reviewing can be a highly time-consuming task, especially when one attempts to be exhaustive in one's overview of the literature. To make sure that no relevant publications are overlooked, large numbers of publications need to be checked, often by going through the reference lists of publications that have already been identified as being relevant. Or the other way around, relevant publications need to be identified by checking all publications that cite one or more publications already identified as being relevant. Bibliographic databases such as Web of Science and Scopus can be used for the above tasks, but they offer only limited functionality to support systematic literature search. CitNetExplorer simplifies systematic literature search in various ways, in particular by making it possible to easily select all publications that cite, or are cited by, a given set of publications.

The above examples of applications of CitNetExplorer all focus on citation networks of scientific publications. We note that CitNetExplorer may also be used to study other types of citation networks, in particular citation networks of patents. For examples of analyses of patent citation networks for which a tool such as CitNetExplorer could be useful, we refer to Barberá-Tomás, Jiménez-Sáez, and Castelló-Molina (2011), Mina, Ramlogan, Tampubolon, and Metcalfe (2007), and Verspagen (2007).

The rest of this paper is organized as follows. In Section 2, we introduce the main concepts that one needs to be familiar with when working with CitNetExplorer. In Section 3, we provide a demonstration of CitNetExplorer. We show applications of CitNetExplorer in which we analyze the scientometric literature and the literature on community detection in networks. Some technical details on the construction, visualization, and analysis of citation networks in CitNetExplorer are discussed in Section 4. We conclude the paper in Section 5.

## 2. Main concepts

We start by discussing a number of important concepts related to CitNetExplorer.

### 2.1. Citation network

CitNetExplorer focuses on citation networks of individual publications. Other types of citation networks, for instance citation networks of journals or researchers, cannot be analyzed using CitNetExplorer. Hence, in CitNetExplorer, each node in a citation network represents a publication. Each edge in a citation network represents a citation relation between two publications. Edges are directed. They start at the citing publication and they end at the cited publication. Edges are unweighted. This is because of the binary nature of citation relations. A publication either does or does not cite a certain other publication.

In CitNetExplorer, citation networks must satisfy two constraints. The first constraint is that citation relations are not allowed to point forward in time. For instance, a publication from 2013 is not allowed to cite a publication from 2014. The second constraint is that citation networks must be acyclic. This for instance means that it is not allowed to have both a citation from publication A to publication B and a citation from publication B to publication A. Likewise, it is not allowed to have a citation from publication A to publication B, a citation from publication B to publication C, and a citation from publication C to publication A. In other words, when moving through a citation network by following citation relations from one publication to another, we should never get back again at a publication that we have already visited. In practice, citation networks are not always perfectly acyclic. There may for instance be publications in the same issue of a journal that mutually cite each other. CitNetExplorer solves this problem by checking whether a citation network is acyclic and by removing citation relations that cause the network to have cycles. There are a number of reasons for requiring citation networks to be acyclic in CitNetExplorer. First, it means that citation networks can be visualized in such a way that all citations flow in the same direction. Second, it allows many of the algorithms included in CitNetExplorer to be implemented in a more efficient way. And third, some concepts used in CitNetExplorer, for instance the concept of the transitive reduction of a citation network (see Subsection 2.3), cannot be used at all without requiring citation networks to be acyclic.

**2.2. Publication attributes**

In CitNetExplorer, each publication has the following attributes:

- *Publication year*. The year in which a publication appeared. As we will see, publication years play an important role in the visualization of a citation network.
- *Citation score*. The number of citations of a publication. CitNetExplorer distinguishes between two types of citation scores, referred to as internal and external citation scores. The internal citation score of a publication is the number of citations of the publication within the citation network being analyzed. The external citation score of a publication is the total number of citations of the publication, including citations from publications outside the citation network being analyzed. Suppose for instance that we use CitNetExplorer to analyze the citation network of all publications in the field of physics in the Web of Science database. The internal citation score of a publication then equals the number of citations received from other physics publications in the Web of Science database, while the external citation score equals the total number of citations received from all publications in the database. By default, CitNetExplorer uses internal citation scores.
- *Marked*. A binary attribute that indicates whether a publication is marked or not. Marked publications play an important role in the drill down functionality of CitNetExplorer (see Subsection 2.4). In the visualization of a citation network, marked publications are represented by a square while non-marked publications are represented by a circle.

- *Selected*. A binary attribute that indicates whether a publication is selected or not. Like marked publications, selected publications play an important role in the drill down functionality of CitNetExplorer. In the visualization of a citation network, a red border is used to indicate that a publication is selected.
- *Group*. The group to which a publication is assigned. In CitNetExplorer, each publication can be assigned to a group (group 1, group 2, etc.). This can be done either manually or algorithmically, for instance using a clustering technique (see Subsection 4.3). A publication cannot be assigned to more than one group. In the visualization of a citation network, colors are used to indicate the group to which a publication belongs. Publications that have not been assigned to a group are colored grey.
- *Complete record*. A binary attribute that indicates whether a publication record is complete or not. For some publications, only an incomplete record may be available. This is for instance the case if a publication is not indexed in the bibliographic database that is used and the publication record has been extracted from the references given to the publication in other publications. In the case of a publication with an incomplete record, it is unknown which publications are cited by this publication. Hence, in CitNetExplorer, the publication receives citations from other publications but appears not to give any citations to other publications. A typical example of publications with an incomplete record are books. The coverage of books in bibliographic databases such as Web of Science and Scopus is limited, and therefore the role of books in the citation network of a research field often can be observed only based on the references that other publications, in particular publications in journals, give to books.

In addition to the above attributes, CitNetExplorer also offers some standard bibliographic data (i.e., authors, title, and source) for each publication.

## 2.3. Transitive reduction

When a citation network includes a large number of citation relations, it is often difficult to provide a clear visualization of the network. To deal with this problem, CitNetExplorer offers the option of displaying only a selection of the citation relations in a citation network. Using this option, only citation relations included in the so-called transitive reduction of the citation network are displayed.

To explain the concept of the transitive reduction of a citation network, we introduce the distinction between what we refer to as essential and non-essential citation relations. A citation relation from publication A to publication B is considered essential if, apart from this relation, there are no other paths in the citation network from publication A to publication B. On the other hand, a citation relation from publication A to publication B is considered non-essential if other paths from publication A to publication B do exist. This would for instance be the case if there is also a citation relation from publication A to publication C and from publication C to publication B. The transitive reduction of a citation network is obtained by removing all non-essential citation relations from the network. Removing all non-essential citation relations has the effect of minimizing the number of citation relations while ensuring that for any pair of publications between which there is a path in the original citation network there still is a path in the transitive reduction.

Figure 1 illustrates the concept of the transitive reduction of a citation network. The transitive reduction of the citation network shown in Figure 1(a) is presented in Figure 1(b).

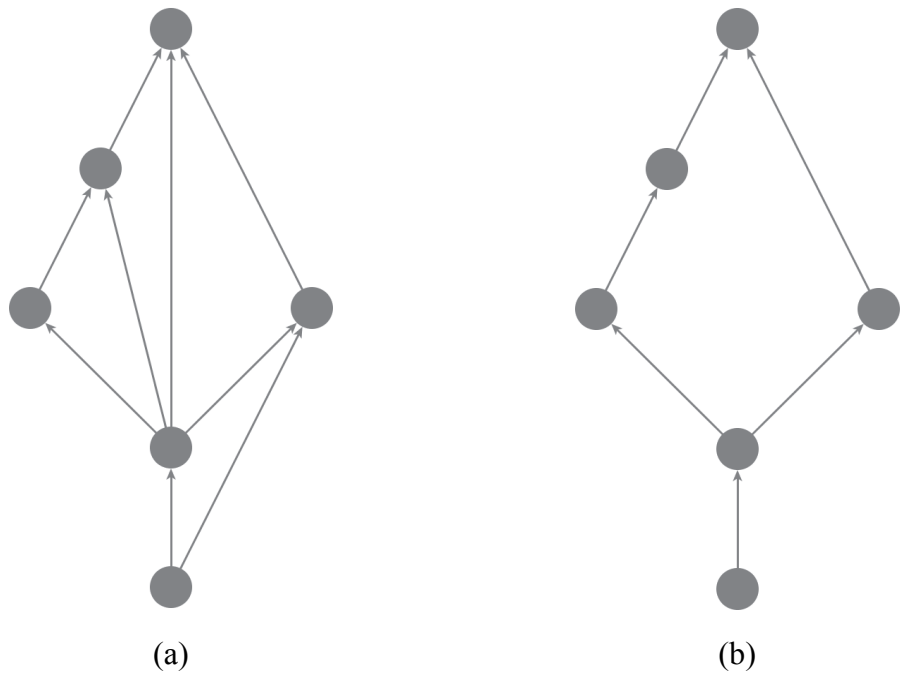

(a)            (b)

Figure 1. Illustration of the concept of the transitive reduction of a citation network. (a) Original citation network. (b) Transitive reduction of the citation network.

We refer to Clough, Gollings, Loach, and Evans (2013) for some further discussion on the transitive reduction of citation networks.

**2.4. Drill down and expansion**

An essential feature for analyzing large citation networks is the functionality offered by CitNetExplorer for drilling down into a citation network. Suppose that we have available the citation network of all publications in the field of scientometrics. Suppose further that we are interested in a specific topic within the field of scientometrics, for instance the topic of the *h*-index. CitNetExplorer provides functionality for drilling down from the citation network of all publications in the field of scientometrics into a subnetwork that includes only a selection of publications, such as all publications on the *h*-index. The original network that we start with is referred to as the full network. The network that is obtained after drilling down is called the current network.

In addition to drill down functionality, CitNetExplorer also offers expansion functionality. Expansion is more or less opposite to drill down. The idea of expansion is to add publications to the current network that, based on citation relations, are closely linked to publications already included in the network. For instance, after drilling down from the citation network of all scientometric publications into a subnetwork of publications on the *h*-index, we may be interested to expand the subnetwork to also include scientometric publications that are not directly about the *h*-index but that are still closely related to this topic. The expansion functionality of CitNetExplorer can be used for this purpose. An illustration of the idea of expansion can be found in an earlier study, where we use the expansion idea to delineate the information science literature (Waltman, Van Eck, & Noyons, 2010).

It is important to mention that multiple consecutive drill downs and expansions can be performed in CitNetExplorer. For instance, starting from the full network, we may perform a drill down. We then obtain a subnetwork, and in this subnetwork we may perform a second drill down. This yields a second subnetwork, smaller than the first one, in which we may perform a third drill down. In the third even smaller subnetwork, we may perform an expansion. We then obtain a fourth subnetwork, larger than the third one, and in this subnetwork we may perform a second expansion.

This results in a fifth subnetwork. It may even be that this fifth subnetwork is in fact the full network from which we started.

To explain the drill down and expansion functionality of CitNetExplorer in more detail, we need to discuss the concepts of predecessors, successors, and intermediate publications. We will first discuss these concepts in the context of the expansion functionality of CitNetExplorer, and we will then discuss them in the drill down context.

*Expansion*

Figure 2(a) shows a simple example of a citation network. Circles represent publications, and arrows represent citation relations. Arrows point in the direction of the cited publication. Suppose that we have already performed a drill down. The publications included in the current network are colored blue.

A predecessor is defined as a publication outside the current network that is cited by one or more publications in the current network. A successor is defined as a publication outside the current network that cites one or more publications in the current network. These definitions are illustrated in Figure 2(b). Predecessors are colored red in this figure, while successors are colored green.[1] The expansion functionality of CitNetExplorer allows us to choose whether we want to expand the current network with predecessors, successors, or both. If we for instance choose to expand the current network with predecessors, the red publications will be added to the current network.

It may be that we do not want to add all predecessors or all successors to the current network but only those predecessors or successors that have a sufficiently strong connection to the publications in the current network. Rather than requiring only a single citation relation with a publication in the current network, we may for instance require three citation relations. This situation is illustrated in Figure 2(c). As can be seen, by requiring three citation relations instead of only one, the number of predecessors or successors that will be added to the current network decreases substantially.

---

[1] Sometimes a publication is both a predecessor and a successor. For simplicity, we do not consider this situation in Figure 2(b).

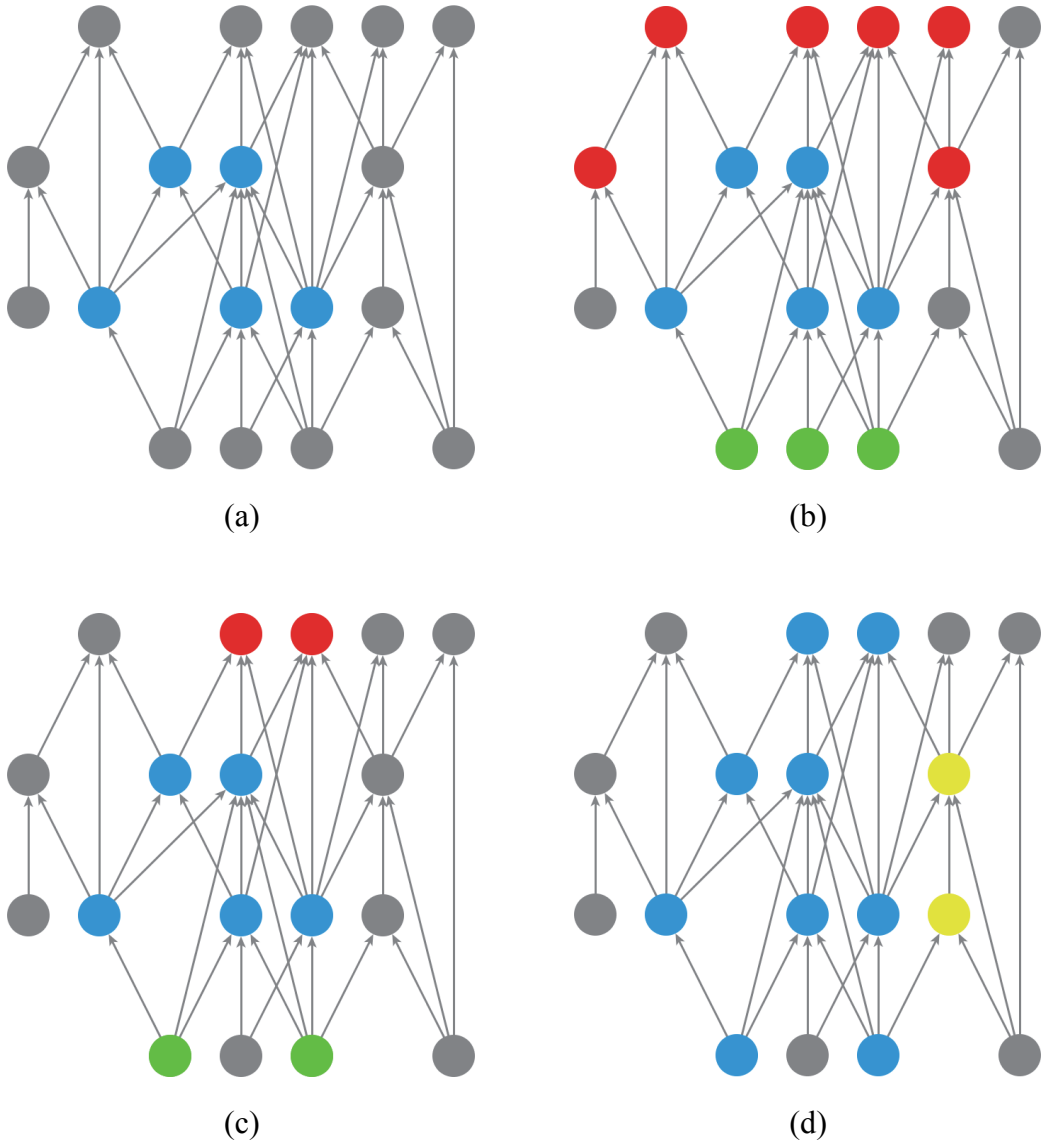

Figure 2. Illustration of the expansion functionality of CitNetExplorer. (a) Publications in the current network are colored blue. (b) Predecessors and successors are colored red and green, respectively. (c) Only predecessors and successors with at least three citation relations with publications in the current network are colored. (d) Intermediate publications are colored yellow.

Suppose that we choose to expand the current network with all predecessors and all successors that have at least three citation relations with publications in the current network. Figure 2(d) shows the situation obtained after the expansion has been performed. The publications included in the current network are colored blue. The publications colored yellow are referred to as intermediate publications. These publications are not included in the current network, but they are located on a citation path between two publications that are included in the current network. In some cases, it may be undesirable that publications located on a citation path between publications included in the current network are not themselves included in the current network. We may not want to have such 'gaps' in the current network. For this reason, the expansion functionality of CitNetExplorer offers the option to expand the current network not only with predecessors and successors but also with intermediate publications. Using this option, the current network would include not only the blue publications but also the yellow ones.

*Drill down*

We now consider the drill down functionality of CitNetExplorer. Drilling down always proceeds in two steps. In the first step, a selection of publications in the current network is made. In the second step, the current network is updated to include only the selected publications. There are three approaches that can be taken to select publications. One approach is to select all publications in a certain time period. Another approach is to select all publications that are assigned to a certain group. The third approach is to mark one or more publications in the current network and to use these marked publications to obtain a selection of publications. This third approach is the default approach in CitNetExplorer. We will now discuss this approach in more detail. As we will see, this approach has a lot in common with the expansion functionality of CitNetExplorer discussed above.

Suppose that we have marked four publications in the current network. This situation is illustrated in Figure 3(a), in which the marked publications are colored blue. In the simplest case, the selected publications coincide with the marked publications. In this case, the blue publications in Figure 3(a) are not only the marked publications but also the selected ones.

However, we may want to select not only the marked publications but also intermediate publications. In the drill down context, intermediate publications are publications in the current network that have not been marked themselves but that are located on a citation path between two marked publications. In Figure 3(b), intermediate publications are colored yellow. If we choose to select not only the marked publications but also the intermediate ones, our selection consists of both the blue and the yellow publications.

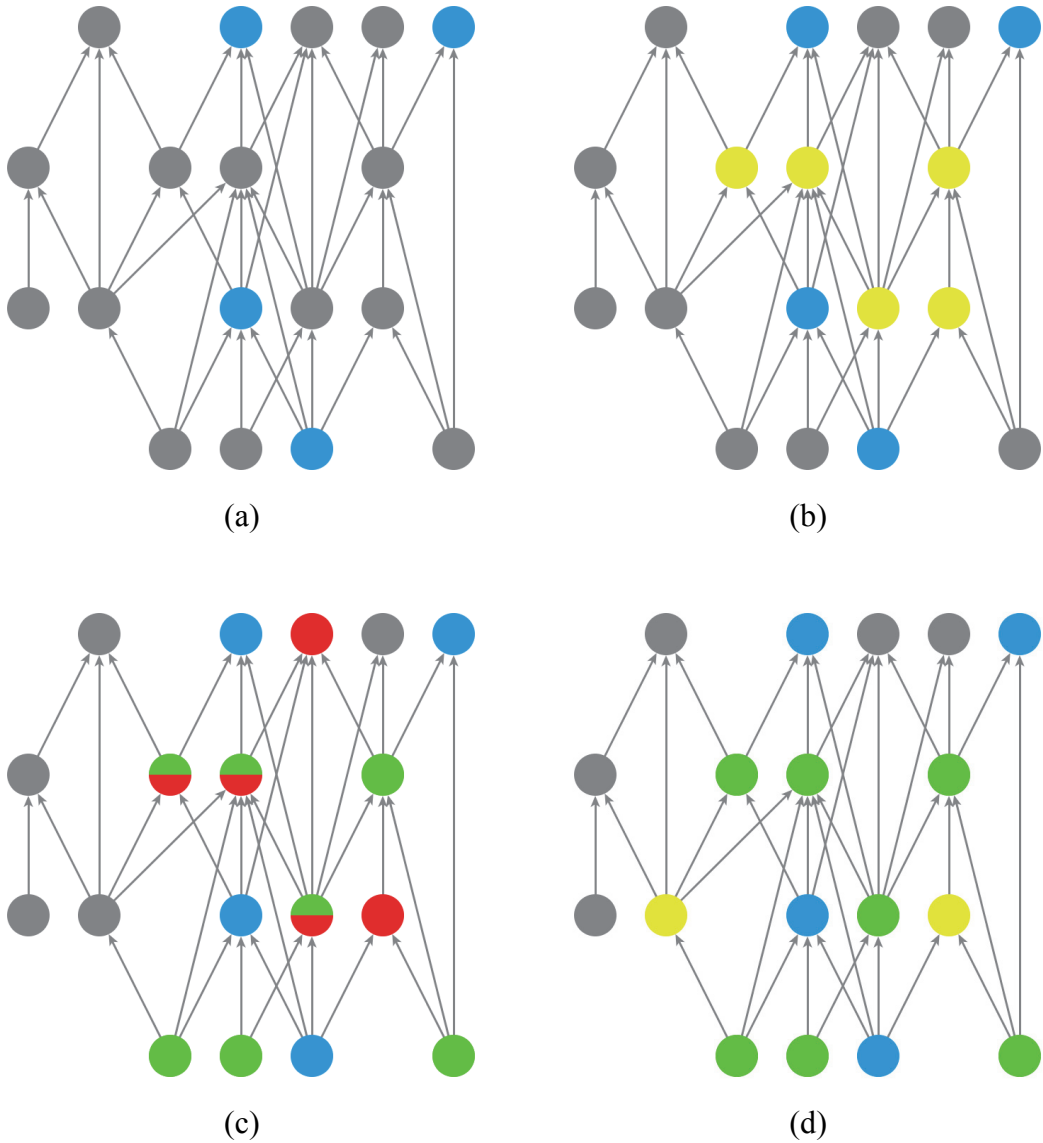

Figure 3. Illustration of the drill down functionality of CitNetExplorer. (a) Marked publications are colored blue. (b) Intermediate publications are colored yellow. (c) Predecessors and successors are colored red and green, respectively. (d) Successors and intermediate publications are colored green and yellow, respectively.

Another possibility is that, in addition to the marked publications themselves, we also want to select predecessors and/or successors of the marked publications. Predecessors are publications in the current network that have not been marked but that are cited by one or more marked publications. Similarly, successors are publications in the current network that have not been marked but that cite one or more marked publications. In Figure 3(c), predecessors are colored red and successors are colored green. Notice that some publications have both a red and a green color.

These publications are both a predecessor and a successor. If we choose to include both predecessors and successors in our selection, the selection consists of all blue, red, and green publications.

A final possibility is that we want our selection to include predecessors and/or successors of the marked publications and also intermediate publications. Suppose for instance that we choose to include successors and intermediate publications in our selection. The selection is then given by all blue, green, and yellow publications in Figure 3(d).

## 3. Demonstration

We now provide a demonstration of CitNetExplorer. We first use CitNetExplorer to analyze the scientometric literature, in particular the literature on the topic of the *h*-index (Hirsch, 2005) and on the topic of science mapping. We then demonstrate an application of CitNetExplorer in which we focus on the topic of community detection in networks (Fortunato, 2010), which is studied in the field of physics. In this application, we work with a very large citation network, including almost two million publications and more than 15 million citation relations.

Our aim in this section is to give a general impression of the possibilities offered by CitNetExplorer. We do not intend to provide detailed instructions on the use of the tool. A tutorial with step-by-step instructions on the use of CitNetExplorer is available online at www.citnetexplorer.nl/gettingstarted/.

### 3.1. User interface

Before showing applications of CitNetExplorer, we first introduce the user interface of the tool. Figure 4 presents a screenshot of the user interface. The user interface is split up in a left and a right part. There also is a menu bar, located in the top part of the user interface.

*Right part*. The right part of the user interface allows the user to switch between two tabs. One tab presents a visualization of the current network. The other tab presents a list of the publications in the current network along with some search functionality. By default, the visualization tab is selected. The interpretation of the visualization is discussed in Subsection 3.2.

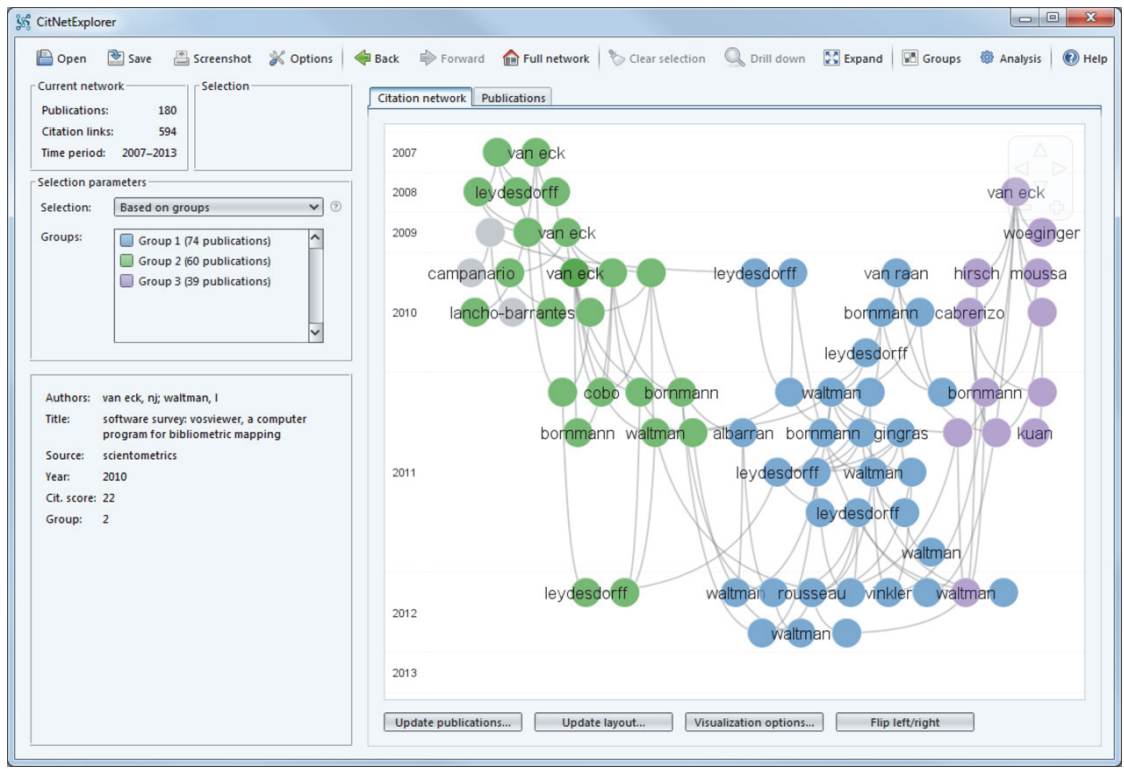

Figure 4. Screenshot of the user interface of CitNetExplorer.

*Left part*. The left part of the user interface serves three purposes. First, it reports the number of publications and the number of citation relations in the current network. If there are selected publications in the current network, it also reports the number of selected publications and the number of citation relations between these publications. Second, the left part of the user interface allows the user to change the way in which publications are selected. As already mentioned in Subsection 2.4, publications can be selected based on their publication year, based on their assignment to a group, or based on their relationship with one or more marked publications. Third, the left part of the user interface is used to present bibliographic data (i.e., authors, title, source, and year) on publications. When in the right part of the user interface the mouse is moved over a publication, bibliographic data on this publication will appear in the left part of the user interface.

*Menu bar*. The menu bar can be used to undertake a number of important actions. For instance, a citation network can be opened or saved, a drill down or expansion can be performed (see Subsection 2.4), or the current network can be analyzed (see Subsection 4.3). Furthermore, in a similar way as we can navigate back and forth between web pages in a web browser, the menu bar can also be used to navigate back and forth between different subnetworks of a citation network.

**3.2. Application to the scientometric literature**

We now demonstrate the use of CitNetExplorer to analyze the scientometric literature.

*Data collection*

Using the Web of Science web interface, bibliographic data on all 25,242 publications in the 13 journals listed in Table 1 was downloaded.[2] To select these journals, we started with *Scientometrics* and *Journal of Informetrics*, the two core scientometric journals. We then used the Journal Citation Reports to identify closely related journals. We took all journals listed among the five most closely related journals to either *Scientometrics* or *Journal of Informetrics*, excluding journals that seem to be mainly nationally oriented. For each selected journal, we also added possible predecessors to the selection. The 25,242 publications in the 13 selected journals relate to the period 1945–2013.

Table 1. Journals included in the data collection.

| Journal | No. of pub. |
|---|---|
| American Documentation | 796 |
| ASLIB Proceedings | 2,697 |
| Information Processing and Management | 3,036 |
| Information Scientist | 254 |
| Information Storage and Retrieval | 372 |
| Journal of Documentation | 3,778 |
| Journal of Information Science | 1,855 |
| Journal of Informetrics | 399 |
| Journal of the American Society for Information Science | 2,995 |
| Journal of the American Society for Information Science and Technology | 2,486 |
| Research Evaluation | 383 |
| Research Policy | 2,596 |
| Scientometrics | 3,595 |

*Analysis: h-index*

After loading the data downloaded from the Web of Science database into CitNetExplorer, we obtain a citation network consisting of 28,482 publications and 158,292 citation relations. The citation network includes 3,240 publications that are not among the 25,242 publications included in the data collection. These are publications that are cited by at least ten publications included in the data collection. Examples are classical scientometric works from the first half of the 20th century, for instance by Bradford and Lotka, but also books, for instance by Garfield, Kuhn, and De Solla Price, and scientometric publications in multidisciplinary journals such as *Nature*, *PLoS ONE*, *PNAS*, and *Science*. We emphasize that for the 3,240 publications not included in the data collection we have only limited data available. For instance, we know the first author of each publication, but we do not know whether there are any additional authors. We also do not know to which publications the 3,240 publications refer.

---
[2] Data collection took place on November 7, 2013.

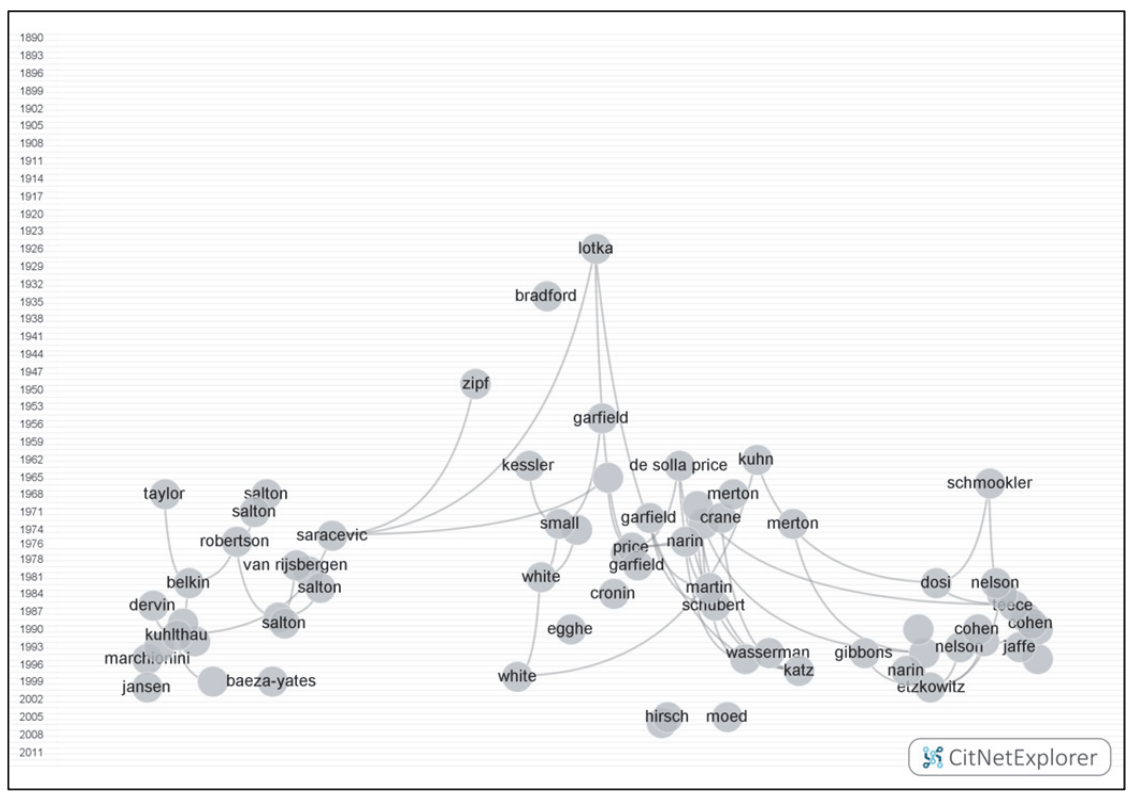

Figure 5. Citation network of the field of scientometrics and closely related fields.

A visualization of our citation network is presented in Figure 5. For obvious reasons, it is not possible to display all 28,482 publications and all 158,292 citation relations in the visualization. In the visualization of a citation network, CitNetExplorer usually includes only a selection of all publications in the network. By default, the 40 most frequently cited publications are included. In the visualization in Figure 5 and also in all other visualizations presented in this subsection, we have chosen to include the 70 most frequently cited publications. In addition, we have also chosen to display only citation relations included in the transitive reduction of the citation network (see Subsection 2.3).

In the visualization of a citation network, CitNetExplorer uses circles to indicate publications (except for marked publications, which are indicated using a square). Curved lines are used to indicate citation relations. The vertical axis in a visualization represents time. The location of a publication in vertical direction is determined by the year in which the publication appeared, with more recent publications being located below older publications. In addition, publications are positioned in vertical direction in such a way that citations always flow upward. So a citing publication is always located somewhere below the corresponding cited publication, even if the two publications appeared in the same year. The locations of publications in horizontal direction are determined by the closeness of publications in the citation network. In general, the closer two publications are to each other in the citation network, the closer to each other they are positioned in horizontal direction.

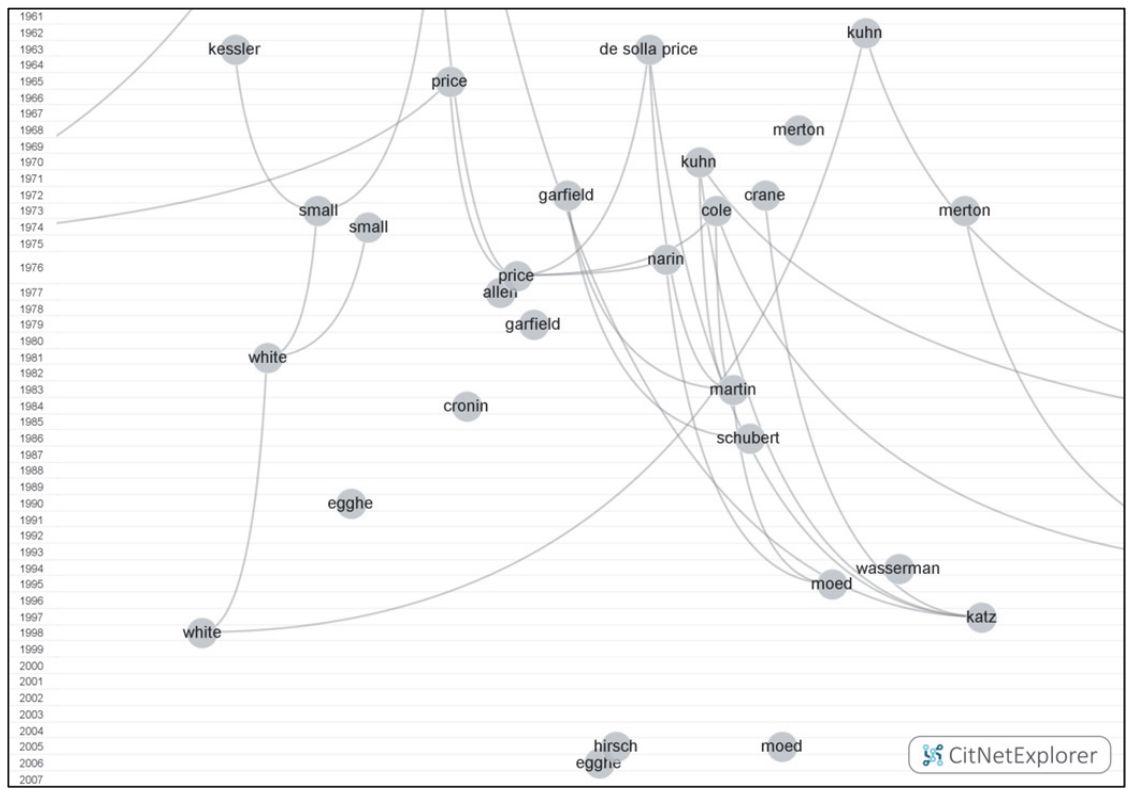

Figure 6. Citation network of the field of scientometrics and closely related fields (zooming in on the central area in the visualization).

By default, in the visualization of a citation network, CitNetExplorer labels publications by the last name of the first author. As can be seen in Figure 5, sometimes the label of a publication is not displayed. This is done in order to avoid labels from overlapping each other. CitNetExplorer offers zooming and panning (scrolling) functionality that can be used to explore a visualization in more detail. By zooming in, labels that initially were not displayed will become visible. This is illustrated in Figure 6, where we have zoomed in on the central area in the visualization in Figure 5.

It is clear that our citation network does not only include scientometric publications. In the left area in the visualization in Figure 5, publications on information science and information retrieval can be found. In the right area, we observe publications on technology and innovation studies. Only the publications in the central area truly represent scientometric research, as can also be seen in Figure 6.

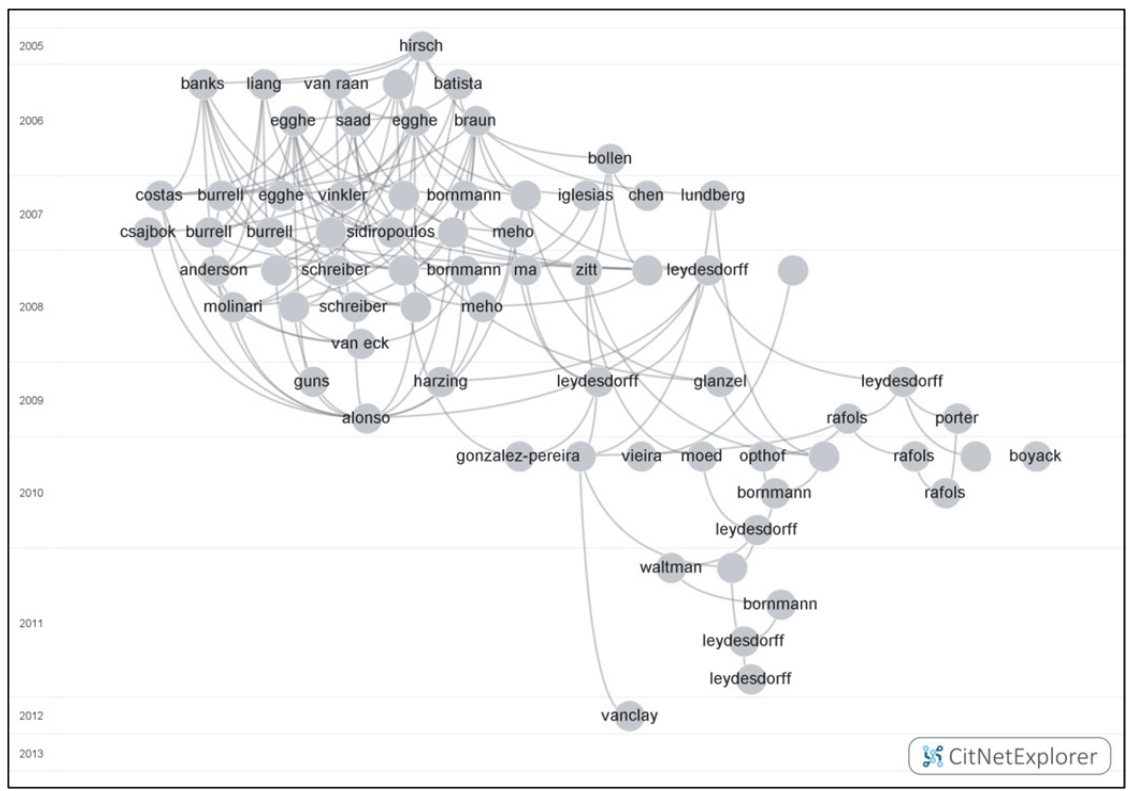

Figure 7. Citation network of the paper by Hirsch in 2005 and its direct and indirect successors.

Since we are interested in studying the literature on the *h*-index, we drill down into the subnetwork consisting of the publication by Hirsch in 2005 in which the *h*- index was introduced and all direct and indirect successors of this publication. Indirect successors are publications that do not themselves cite the publication by Hirsch in 2005 but that are for instance successors of successors or even successors of successors of successors of this publication. So an indirect successor is a publication from which there is a path in the citation network to the publication by Hirsch in 2005, where the path is of length at least two. Figure 7 shows a visualization of the subnetwork. The subnetwork includes 1,371 publications, of which the 70 most frequently cited ones are included in the visualization.

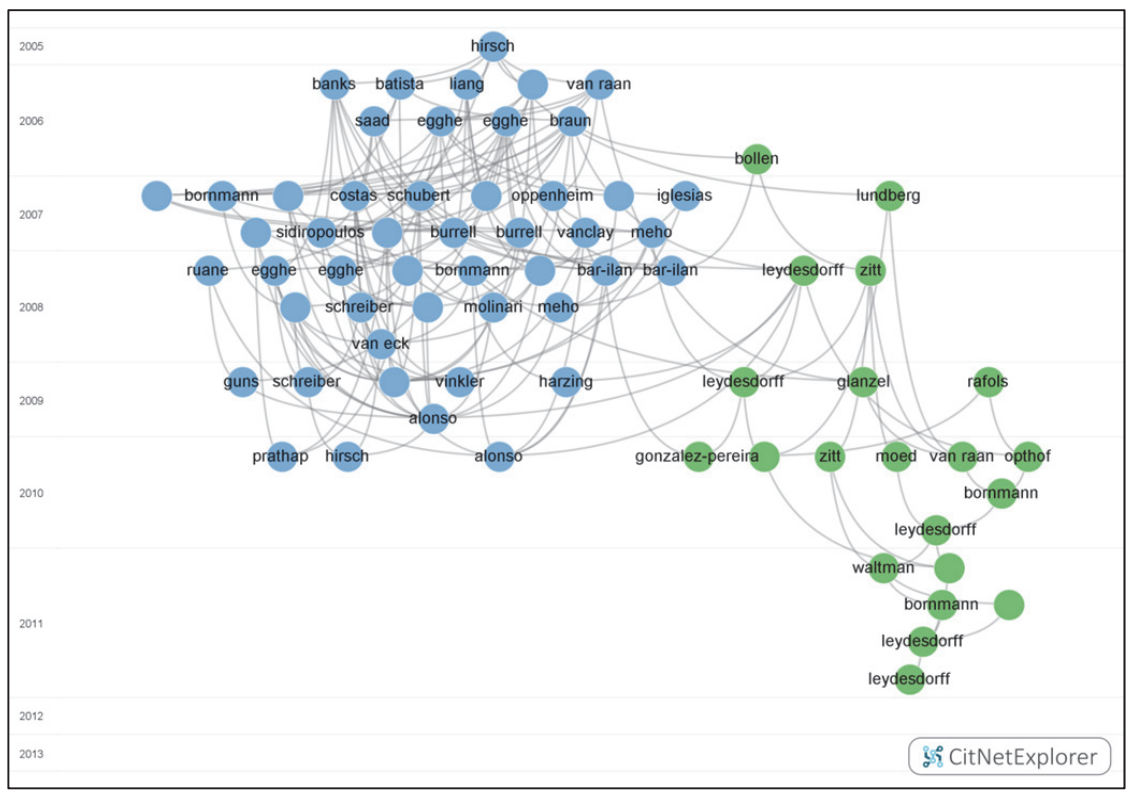

Figure 8. Citation network of the paper by Hirsch in 2005 and the core of its direct and indirect successors.

Many publications that (directly or indirectly) cite the publication by Hirsch in 2005 may be only weakly related to the topic of the *h*-index. Among the 1,371 publications in our subnetwork, we therefore make a selection of core publications (see also Subsection 4.3). We define a core publication as a publication that has citation relations with at least ten other core publications. Based on this criterion, 230 core publications are identified. We drill down into the subnetwork consisting of these 230 publications. After drilling down, we cluster the publications (see also Subsection 4.3). Two clusters are identified, each consisting of publications that are strongly connected to each other in terms of citation relations. Figure 8 shows the visualization that we obtain after clustering the publications. Based on our knowledge of the scientometric literature, it is immediately clear that the blue cluster, located in the left area in the visualization, consists of publications on the *h*-index and its variants. This cluster includes 174 publications. The remaining 56 publications can be found in the green cluster, located in the right area in the visualization. These publications are not directly about the *h*-index but instead deal with the closely related topic of advanced citation-based indicators.

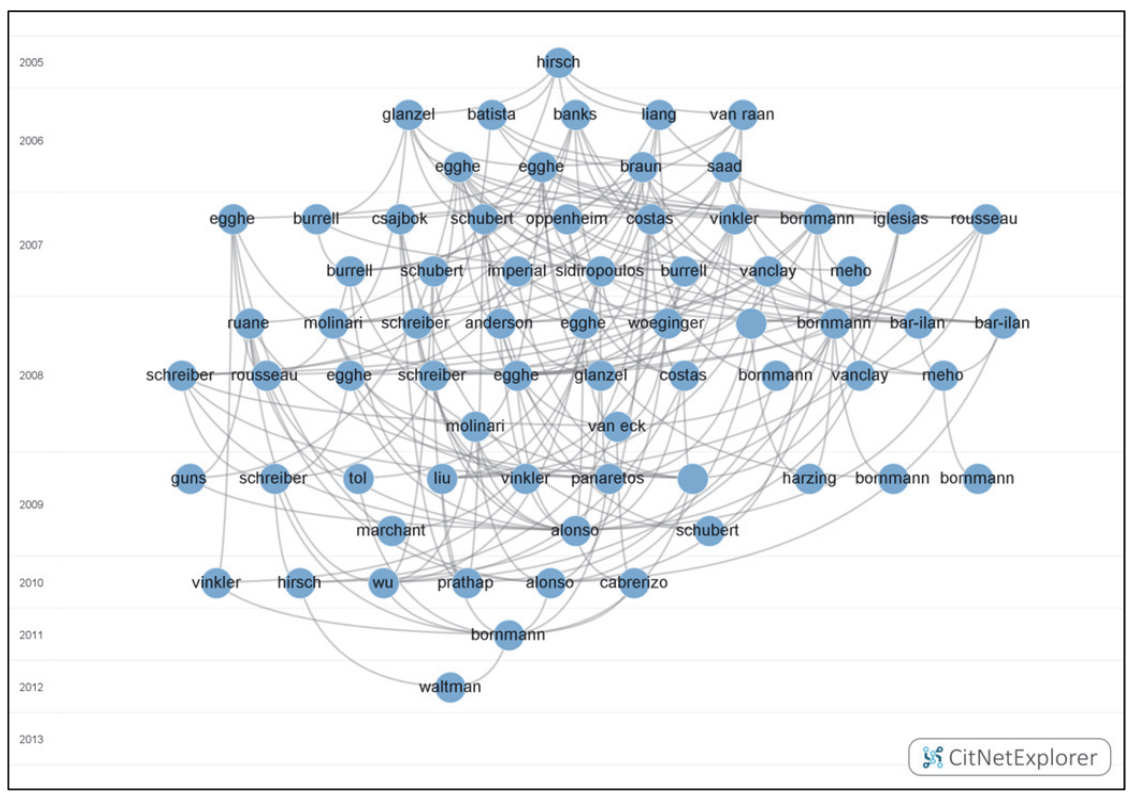

Figure 9. Citation network of the literature on the *h*-index and its variants.

Drilling down into the blue cluster yields the visualization presented in Figure 9. The visualization displays the citation network of the most frequently cited publications on the *h*-index, starting with the publication by Hirsch in 2005 and ending with the publication by Waltman on the inconsistency of the *h*-index in 2012. Based on Figure 9, we clearly see that the *h*-index literature constitutes a very dense citation network. Because of the denseness of the citation network, CitNetExplorer turns out to be unable to identify a more detailed structure in the *h*-index literature.[3]

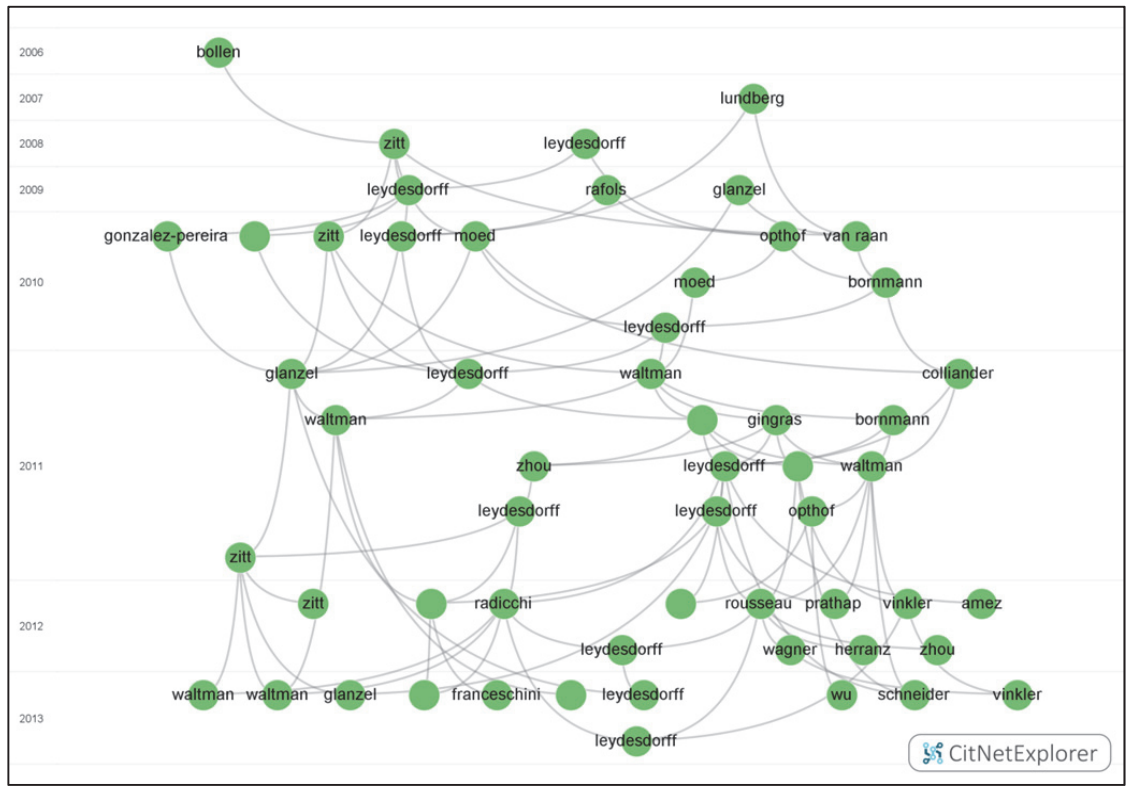

Figure 10. Citation network of the literature on advanced citation-based indicators.

As already mentioned in Subsection 3.1, CitNetExplorer allows us to navigate back and forth between different subnetworks of a citation network. After moving back to our subnetwork consisting of 230 publications, we drill down into the green cluster. Figure 10 shows the visualization that we obtain. Since the green cluster includes only 56 publications, all publications are displayed in the visualization. The visualization gives an overview of the development of the literature on advanced citation-based indicators after the introduction of the *h*-index, starting with well-known publications by Bollen, Lundberg, and Zitt and ending with recent work by for instance Glänzel, Leydesdorff, and Waltman. The visualization displays the structure of the literature in a quite detailed way. In the right area in the visualization, we find publications dealing with indicators based on the principle of cited-side normalization. In the left area, on the other hand, we see publications on indicators based on the citing-side normalization principle and also publications on recursive variants of these indicators (inspired by the PageRank algorithm).

---

[3] We refer to Zhang, Thijs, and Glänzel (2011) for an alternative approach to analyzing the *h*-index literature. This approach seems to give a somewhat more detailed picture of the structure of the literature.

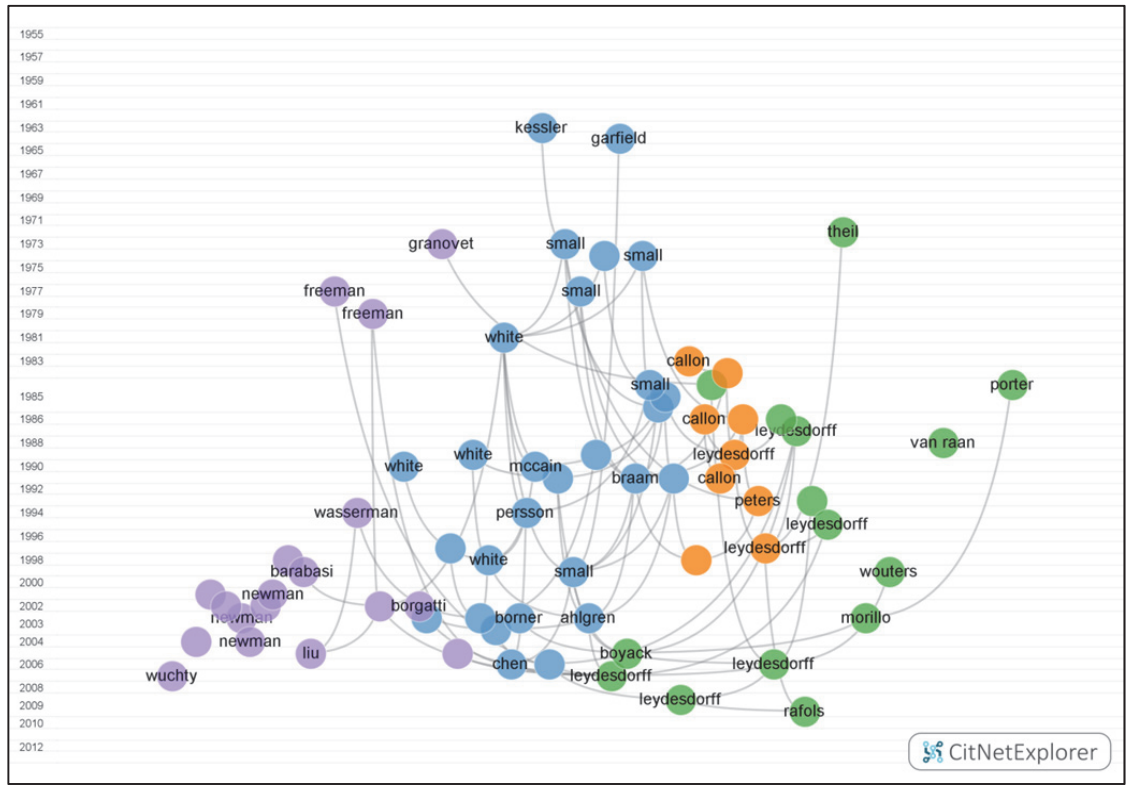

Figure 11. Citation network of the literature on science mapping.

*Analysis: Science mapping*

We now look at the scientometric literature on the topic of science mapping. We first move back to the full network, and we then cluster the 28,482 publications in this network. Because we want to obtain a clustering at a relatively high level of detail, we use a value of 5.00 for the resolution parameter instead of the default value of 1.00. The resolution parameter is the most important parameter of the clustering technique used by CitNetExplorer. As we explain more extensively in Subsection 4.3, the resolution parameter determines the level of detail at which clusters are identified. The higher the value of the parameter, the larger the number of clusters that will be obtained. Using a value of 5.00 for the resolution parameter, it turns out that publications by Small, Wasserman, and White are assigned to the same cluster, suggesting that this cluster covers the topic of science mapping. We drill down into the subnetwork consisting of the 1,105 publications in our science mapping cluster.

We then create another clustering, but this clustering involves only the 1,105 publications in our subnetwork. This time we use the default value of 1.00 for the resolution parameter. Figure 11 shows the visualization that is obtained.

We observe four clusters in the visualization in Figure 11. The blue cluster can be considered to cover the core of the science mapping literature, in particular the work on co-citation and bibliographic coupling analysis, the orange cluster mainly covers the topic of co-word analysis, and the purple cluster covers the topic of (social) network analysis. The green cluster is more difficult to label. On the one hand it seems to cover the topic of interdisciplinarity, but on the other hand it also includes a large number of publications from the same author. Of the 248 publications in the green cluster, 57 are authored by Leydesdorff. This suggests that to some degree the cluster may represent the oeuvre of an author rather than a scientific topic.

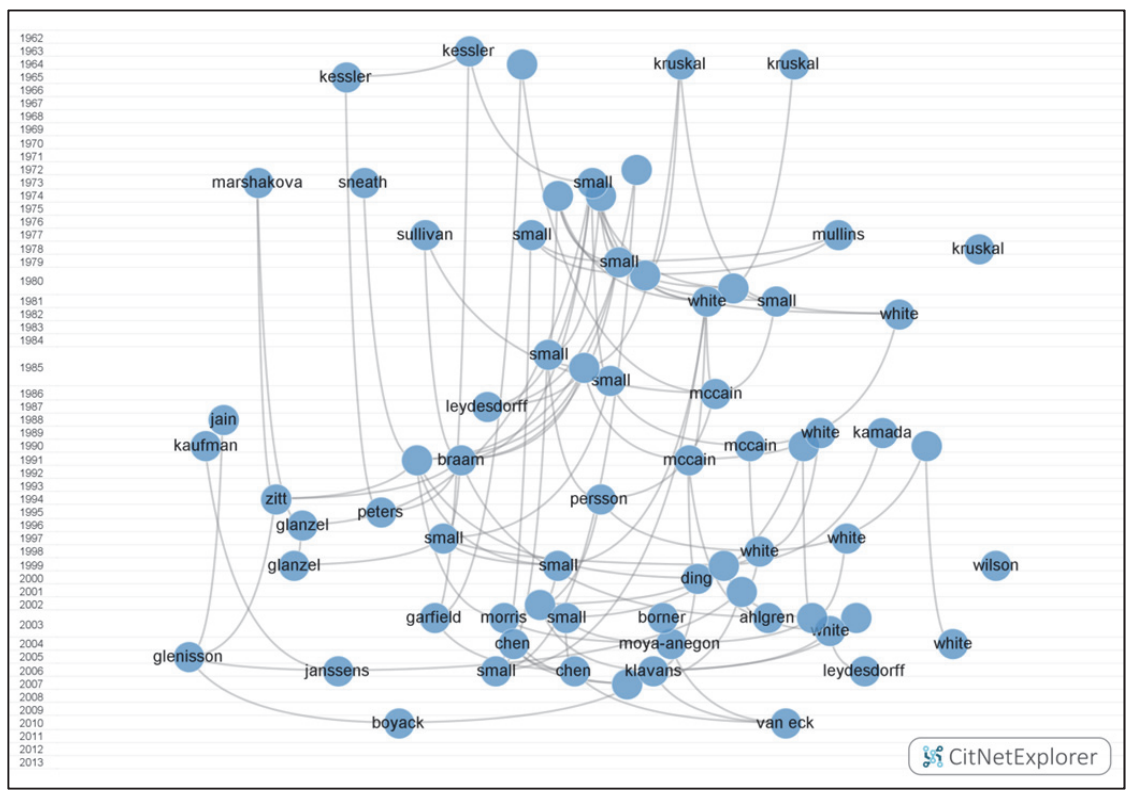

Figure 12. Citation network of the literature on co-citation and bibliographic coupling analysis.

We now drill down into the blue cluster. This yields the visualization presented in Figure 12. Many well-known publications related to science mapping can be found in the visualization, for instance classical publications by Kessler (bibliographic coupling analysis), Kruskal (multidimensional scaling), and Small (co-citation analysis), more recent publications by, among others, Börner, Boyack, Chen, Klavans, and Van Eck, and also, in the right area in the visualization, a substantial number of publications on author co-citation analysis.

CitNetExplorer can also be used to study citation paths through the scientific literature. This is somewhat similar to the idea of 'document pathways' suggested by Small (1999). To illustrate the analysis of citation paths, we mark the publication by Kessler in 1963, in which the concept of bibliographic coupling was introduced, and the publication by Van Eck in 2010, in which the VOSviewer software was presented. We then identify the longest path in the citation network between the two marked publications. There turn out to be multiple longest paths, all of length 21. These paths are shown in Figure 13. Most publications on these paths deal with the topic of author co-citation analysis. We may also be interested in the shortest rather than the longest path between two publications. In the case of our two marked publications, this results in a large number of paths (not shown) that are all of length three.

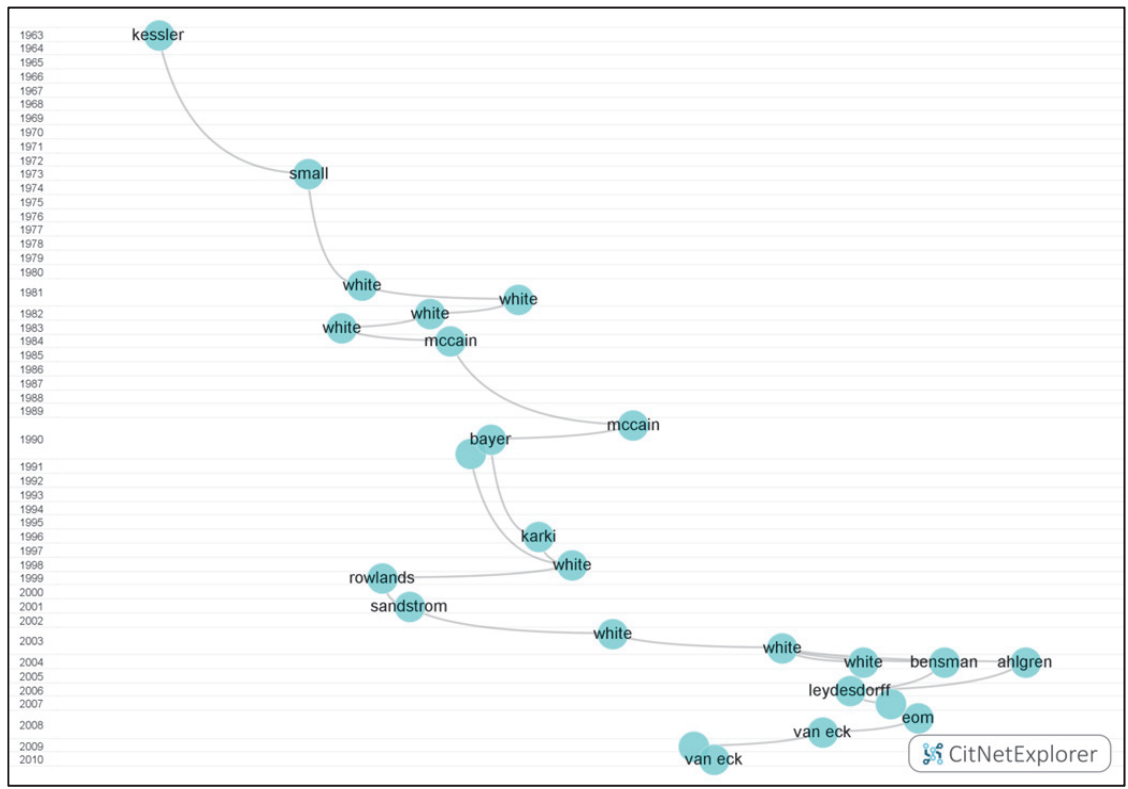

Figure 13. Longest citation paths between the publication by Kessler in 1963 and the publication by Van Eck in 2010.

**3.3. Application to the literature on community detection in networks**

We proceed by considering an application of CitNetExplorer in which we study the literature on the topic of community detection in networks. This topic has received a lot of attention during the past 10 to 15 years. Most research has appeared in journals in the field of physics. We refer to Fortunato (2010) for a review of the community detection literature. Compared with the above-discussed application of CitNetExplorer to the scientometric literature, we now start with a much larger set of publications. We also use a different strategy to search for relevant publications.

*Data collection*

Bibliographic data on all publications in physics journals in the Web of Science database in the period 1998–2012 was collected. In addition to physics journals, multidisciplinary journals such as *Nature* and *Science* were also included in the data collection. Downloading data on such a very large number of publications using the Web of Science web interface is extremely inconvenient. Instead, we obtained the data from the in-house version of the Web of Science database of our institute. Data was collected on about 1.8 million publications and on about 15.1 million citation relations between these publications.

*Analysis*

Our aim is to get an overview of the development of the literature on community detection in networks. As we will see, only a very small share of the 1.8 million publications included in our citation network deal with the topic of community detection. To identify these publications, we first use the search functionality of CitNetExplorer, followed by the expansion functionality. This is different from the analysis of the scientometric literature presented in Subsection 3.2, where for instance the clustering functionality of CitNetExplorer plays an important role. The idea of first searching for publications and then expanding the search results based on citation relations has been employed in earlier studies, some of them with an information retrieval focus (e.g., Cawkell, 1974; Larsen, 2002) and others with a scientometric focus (e.g., Zitt & Bassecoulard, 2006). Using CitNetExplorer, identifying publications in this way can be done without much effort. Also, the effect of different search criteria or different expansion parameters can be easily tested.

Figure 14 shows a visualization of our citation network. In this visualization and also in the other visualizations presented in this subsection, only the 40 most frequently cited publications are included. Furthermore, only citation relations included in the transitive reduction of the citation network (see Subsection 2.3) are displayed. In the top-left area in the visualization, we observe publications on various topics in solid-state physics. In the bottom-left area, we find publications on graphene. Publications on complex networks can be found in the central area in the visualization. Finally, publications on particle physics and string theory are located in the right area.

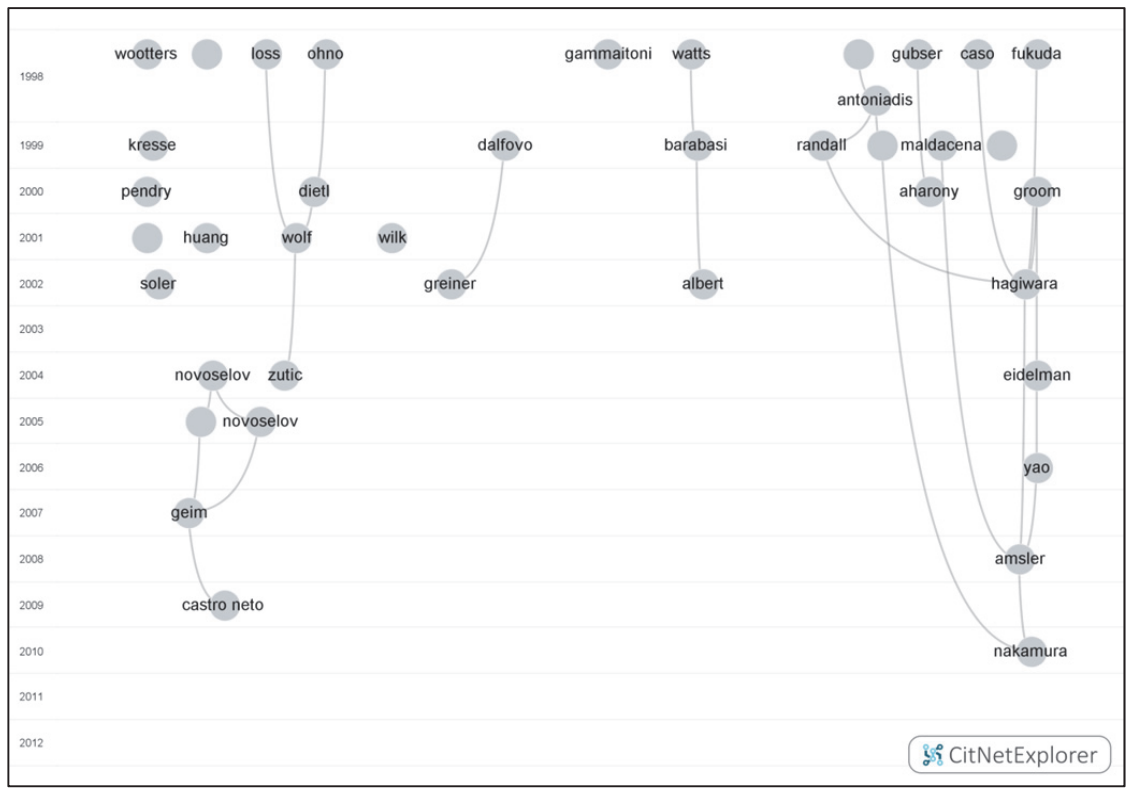

Figure 14. Citation network of the field of physics.

Among the 1.8 million publications in our citation network, we search for publications whose title matches '*communit* detect*' or '*detect* communit*'. There turn out to be 113 publications with a matching title. We drill down into the subnetwork consisting of these 113 publications. Some of the 113 publications are false positives and do not deal with community detection. An example is a publication with the title 'Detection of large numbers of novel sequences in the metatranscriptomes of complex marine microbial communities'. We assume that false positives have no citation relations with publications that have been correctly identified as being relevant to the topic of community detection. We therefore identify the largest connected component in our subnetwork and remove from the subnetwork the seven publications not included in this component. We now have a subnetwork that includes 106 publications. A visualization of this subnetwork is presented in Figure 15.

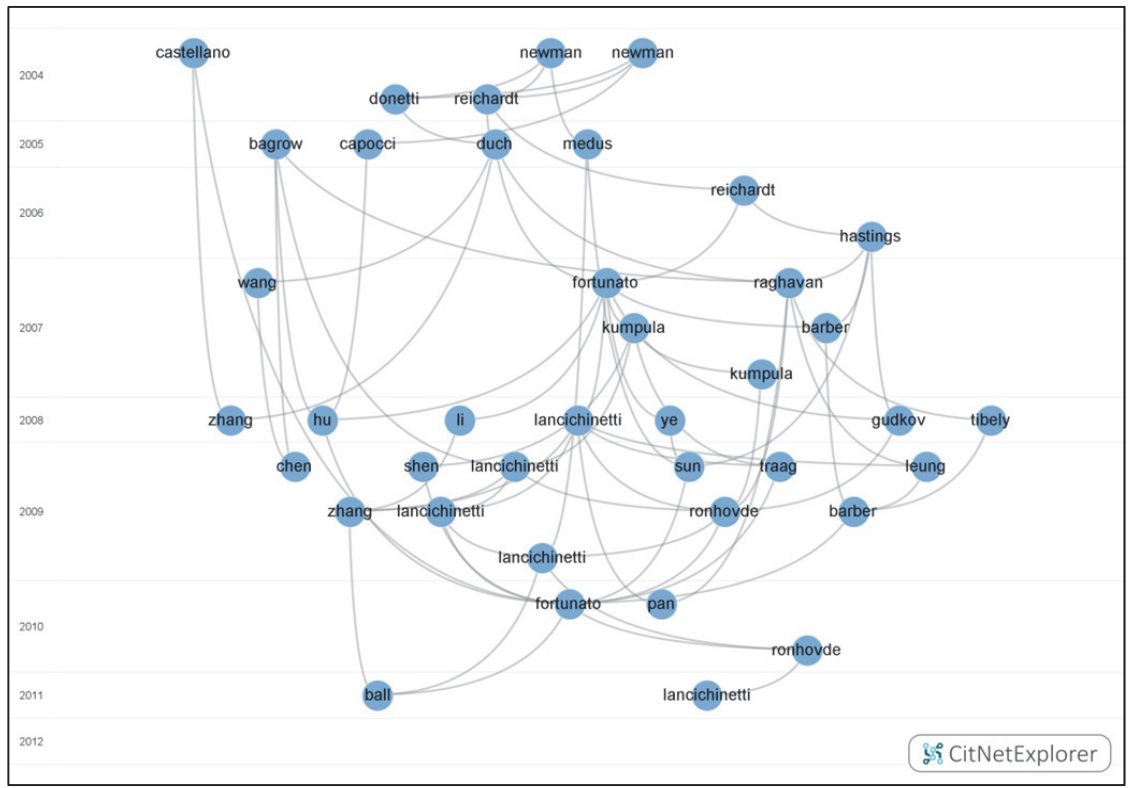

Figure 15. Preliminary citation network of the literature on community detection.

Based on our knowledge of the community detection literature, we observe that some important early publications on community detection are not included in our subnetwork. We therefore expand the subnetwork with 33 predecessors that are each cited by at least ten publications already included in the subnetwork. In the resulting subnetwork of 139 publications, some of the early publications turn out to deal with networks but not specifically with community detection. An example is a well-known review article with the title 'Statistical mechanics of complex networks'. We manually identify six publications not dealing specifically with community detection, and we order CitNetExplorer to remove these publications from the subnetwork.

Finally, we expand our subnetwork of 133 publications with 439 successors that each cite at least four publications already included in the subnetwork. Figure 16 shows a visualization of the subnetwork that is obtained after the expansion. The visualization offers a clear picture of the development of the community detection literature. Based on our knowledge of the literature, we interpret the visualization as follows. The problem of community detection in networks was raised in a publication by Girvan in 2002. In 2004, a number of important publications on community detection appeared, especially the work by Newman on modularity-based community detection. In the next years, researchers proposed algorithms for modularity-based community detection (Blondel, Duch, and Newman), they studied issues related to resolution parameters (Reichardt) and resolution limits (Fortunato), and they introduced alternative approaches to community detection (Rosvall), some of which allow for communities that are overlapping (Lancichinetti and Palla) or hierarchically organized (Arenas, Clauset, Lancichinetti, and Sales-Pardo). Researchers also started to pay attention to the issue of comparing different community detection approaches (Danon and Lancichinetti). In 2010, the community detection literature was summarized in a review article by Fortunato.

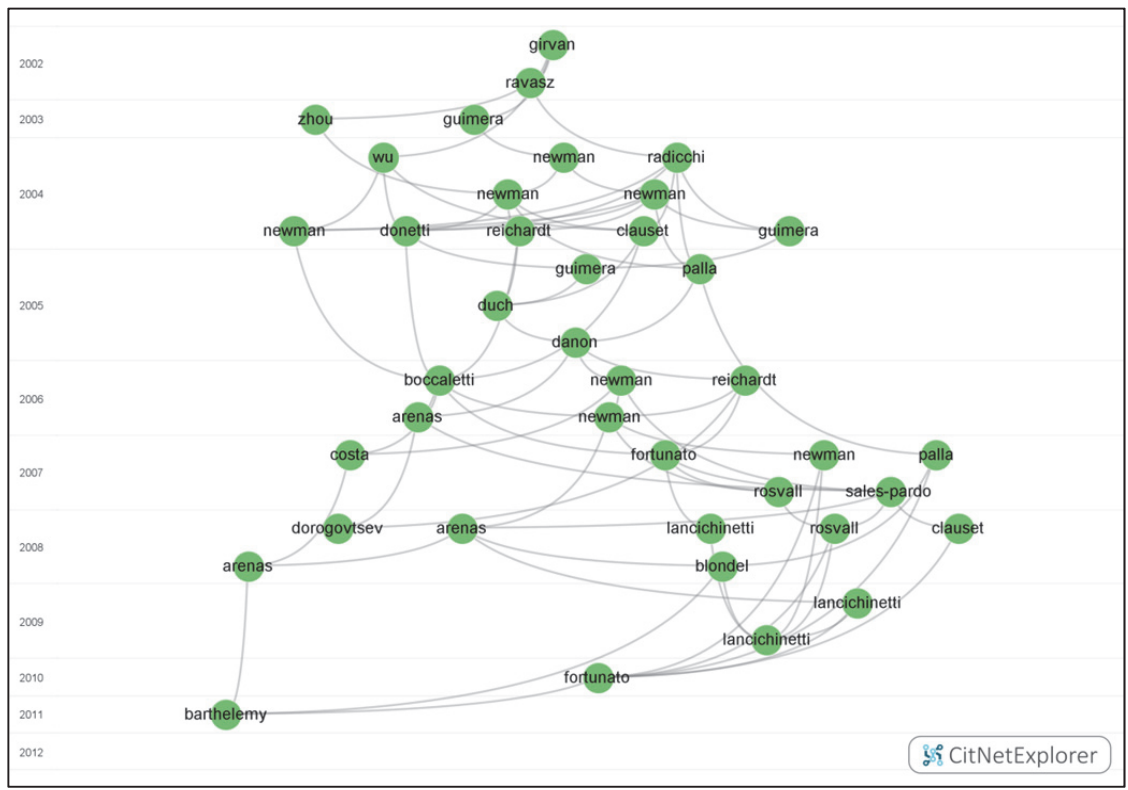

Figure 16. Final citation network of the literature on community detection.

Although we believe that the visualization in Figure 16 provides an accurate overview of the development of the community detection literature, we also need to mention two limitations of the visualization. One limitation is that the most recent years are underrepresented. This is a consequence of the fact that recent publications tend to have fewer citations than older publications. Recent publications are therefore less likely to be among the 40 most frequently cited publications that are included in the visualization. The second limitation is that the visualization includes a number of a review articles that are not primarily about community detection. These publications have been added in the final expansion step described above. Review articles tend to have many citation relations (both incoming and outgoing) with other publications, and they therefore have a high likelihood of being added to a subnetwork when CitNetExplorer's expansion functionality is used. This may lead to the inclusion of review articles that are only weakly related to the literature being analyzed.

## 4. Technical details

In this section, technical details on the construction, visualization, and analysis of citation networks in CitNetExplorer are discussed.

### 4.1. Construction of citation networks

CitNetExplorer requires data based on which a citation network can be constructed. There are two approaches that can be taken. One approach is to provide CitNetExplorer both with data on publications and with data on citation relations between publications. This approach is taken in Subsection 3.3. When this approach is used, the construction of a citation network is straightforward, since all required data is provided directly to CitNetExplorer. The other approach, taken in Subsection 3.2, is to provide CitNetExplorer with data downloaded using the Web of Science web interface. In this approach, CitNetExplorer is provided with data on publications, but it does not directly receive data on citation relations between publications. Instead, based on the cited reference data of publications, CitNetExplorer needs to find out which publication cites which other publication. To do so, CitNetExplorer attempts to match each cited reference with a publication. This is known as citation matching.

Citation matching can be performed in different ways. In CitNetExplorer, citation matching is performed as follows. CitNetExplorer first tries to match based on DOI. However, DOI data often is not available. In that case, matching is done based on first author name (last name and first initial only), publication year, volume number, and page number. A perfect match is required for each of these data elements. Data on the title of the cited journal usually is available as well, but this data is not used by CitNetExplorer. This is because in many cases the title of a journal is not written in a consistent way, making it difficult to perform accurate matching based on journal title.

As discussed in Subsection 2.1, a citation network must satisfy two constraints in CitNetExplorer. One constraint is that citation relations are not allowed to point forward in time. The other constraint is that a citation network must be acyclic. The final step in the construction of a citation network is to remove citation relations that cause these constraints to be violated.

**4.2. Visualization of citation networks**

To create a visualization of a citation network, CitNetExplorer needs to assign each publication in the network to a location in the horizontal and vertical dimensions of the visualization. This is known as a hierarchical graph drawing problem. Following the literature on hierarchical graph drawing (Healy & Nikolov, 2013), CitNetExplorer first assigns each publication to a location in the vertical dimension. As discussed in Section 3, the vertical dimension of the visualization of a citation network represents time. The location of a publication in this dimension depends on the year in which the publication appeared. After for each publication a location in the vertical dimension has been determined, publications are assigned to a location in the horizontal dimension. Below, we first discuss the positioning of publications in the vertical dimension. We then consider the horizontal positioning of publications.

*Positioning publications in the vertical dimension*

The vertical dimension of the visualization of a citation network consists of layers. Each year is represented by one or more layers. Publications are assigned to layers on a year-by-year basis. If the publications from a given year do not cite each other, they are all assigned to the same layer, unless the number of publications exceeds the maximum number of publications per layer. By default, the maximum number of publications per layer is 10. If the publications from a given year do cite each other, they are assigned to multiple layers in such a way that the layer of a citing publication is always located below the layer of the corresponding cited publication. This ensures that citations always flow upward. Assigning publications to multiple layers is accomplished using a simple heuristic algorithm that aims to minimize the number of layers that are needed.

*Positioning publications in the horizontal dimension*

We now consider the positioning of publications in the horizontal dimension of the visualization of a citation network. In the horizontal dimension, publications are positioned based on their closeness to each other in the citation network. Publications that are close to each other in the citation network tend to be positioned close to each other in the horizontal dimension.

The locations of publications in the horizontal dimension are determined by minimizing

$$Q_{\text{visualization}}(x_1,\ldots,x_n) = \sum_{i=1}^{n}\sum_{j=1}^{n}\left(s_{ij}d_{ij}^2 - \alpha d_{ij}^\beta\right), \quad (1)$$

where *n* denotes the number of publications included in the visualization of the citation network, $s_{ij}$ denotes the closeness of publications *i* and *j* in the citation network, *α* denotes the weight of the repulsive force, *β* denotes the exponent of the repulsive force, $x_i$ denotes the location of publication *i* in the horizontal dimension, and $d_{ij}$ denotes the distance between publications *i* and *j* in the horizontal dimension, that is,

$$d_{ij} = |x_i - x_j|. \quad (2)$$

The closeness of publications *i* and *j* in the citation network is calculated based on the idea of making a random walk in the citation network. The details of the calculation are somewhat complex, but essentially the closeness of publications *i* and *j* equals the probability that a random walk starting from publication *i* will end at publication *j*. In this random walk idea, the citation network is treated as undirected. The values of the parameters *α* and *β* in (1) can be chosen by the user. The default values of these parameters are 0.1 and 0.5, respectively. The parameters can be used to fine-tune the horizontal positioning of publications. A more in-depth discussion of (1) is beyond the scope of this paper. However, (1) is closely related to the VOS mapping technique. We refer to Van Eck, Waltman, Dekker, and Van den Berg (2010) for more details on this technique.

Minimizing (1) is done using a simple heuristic algorithm. This algorithm treats the location of a publication in the horizontal dimension as a discrete variable. For each publication *i*, we have

$$x_i \in \left\{ \frac{0}{m-1}, \frac{1}{m-1}, \ldots, \frac{m-1}{m-1} \right\},  \qquad (3)$$

where *m* denotes the number of grid points. The default value of the parameter *m* is 100. There is an additional parameter that determines the minimum distance between publications that have been assigned to the same layer in the vertical dimension. By default, the distance between publications in the same layer must be at least 5 grid points. Eq. (1) is minimized subject to this constraint.

**4.3. Analysis of citation networks**

CitNetExplorer offers a number of options for analyzing citation networks. The following options are available:

- Extracting connected components.
- Clustering publications.
- Identifying core publications.
- Finding the shortest or the longest path from one publication to another.

All four options are demonstrated in Section 3. The first three options ignore the direction of citations. They treat a citation network as if it is undirected. The fourth option does take into account the direction of citations. Below, we provide some more details on the second and the third option.

*Clustering publications*

Clustering publications means that each publication in a citation network is assigned to a cluster in such a way that publications which are close to each other in the citation network tend to be in the same cluster. So each cluster consists of publications that are strongly connected to each other in terms of citation relations. A cluster can usually be interpreted to represent a topic in the scientific literature.

The clustering technique used by CitNetExplorer is documented in an earlier paper (Waltman & Van Eck, 2012). The technique uses a variant of the modularity function of Newman and Girvan (2004) and Newman (2004). More specifically, publications are assigned to clusters by maximizing

$$Q_{\text{clustering}}(u_1,\ldots,u_n) = \sum_{i=1}^{n}\sum_{j=1}^{n} \delta(u_i, u_j)\left(a_{ij} - \frac{\gamma}{2n}\right), \qquad (4)$$

where $n$ denotes the number of publications in the citation network, $a_{ij}$ denotes the relatedness of publications $i$ and $j$, $\gamma$ denotes a so-called resolution parameter, and $u_i$ denotes the cluster to which publication $i$ is assigned. The function $\delta(u_i, u_j)$ equals 1 if $u_i = u_j$ and 0 otherwise. The relatedness of publications $i$ and $j$ is given by

$$a_{ij} = \frac{c_{ij}}{\sum_{k=1}^{n} c_{ik}}, \qquad (5)$$

where $c_{ij}$ equals 1 if publication $i$ cites publication $j$ and 0 otherwise. Hence, if publication $i$ cites publication $j$, the relatedness of the publications is inversely proportional to the total number of publications in the citation network that are cited by publication $i$. If publication $i$ does not cite publication $j$, the relatedness of the publications equals 0. The value of the resolution parameter $\gamma$ in (4) is chosen by the user. The higher the value of this parameter, the larger the number of clusters that will be obtained. The default value of the resolution parameter is 1. We refer to Waltman and Van Eck (2012) for a more extensive discussion of the clustering technique used by CitNetExplorer.

In order to maximize (4), CitNetExplorer uses the smart local moving algorithm that we introduced in a recent paper (Waltman & Van Eck, 2013). This algorithm can be seen as a more sophisticated alternative to the well-known Louvain algorithm for modularity optimization (Blondel, Guillaume, Lambiotte, & Lefebvre, 2008).

CitNetExplorer's clustering technique usually identifies a relatively limited number of larger clusters and a more substantial number of smaller clusters. Sometimes clusters are very small and for instance include only one or two publications. Because in many cases small clusters are of limited interest, CitNetExplorer allows the user to specify a minimum cluster size. Clusters that are too small can be either discarded or merged with other clusters.

*Identifying core publications*

Identifying core publications is about identifying publications that can be considered to constitute the core of a citation network. This option can be used to get rid of unimportant publications in the periphery of a citation network. The identification of core publications is based on the concept of *k*-cores introduced by Seidman (1983). A core publication is defined as a publication that has at least a certain minimum number of citation relations with other core publications. These can be either incoming or outgoing citation relations. The minimum number of citation relations is chosen by the user. The higher the value of this parameter, the smaller the number of core publications. Figure 17 illustrates the concept of core publications.

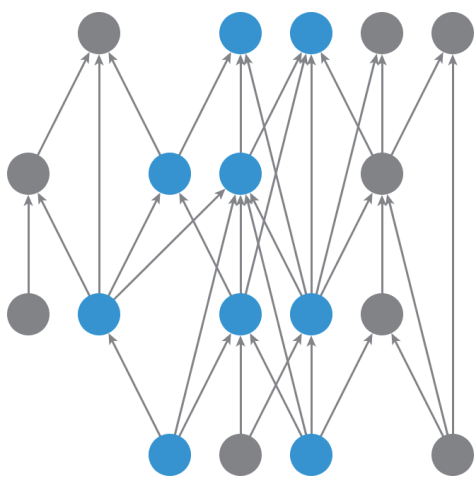

Figure 17. Illustration of the concept of core publications. In this illustration, a core publication is defined as a publication that has at least three citation relations with other core publications. Core publications are colored blue.

## 5. Conclusion

We have introduced CitNetExplorer, a new software tool for analyzing and visualizing citation networks. The most important concepts that need to be understood when working with CitNetExplorer have been presented. Also, CitNetExplorer has been demonstrated by showing two applications, one in which the scientometric literature is studied and another one focusing on the literature on community detection in networks. Finally, we have discussed some technical details on the construction, visualization, and analysis of citation networks in CitNetExplorer.

The scientometric community has focused more on the analysis of co-citation and bibliographic coupling networks than on the analysis of direct citation networks. The community is therefore still relatively inexperienced in the types of analyses made possible by a tool such as CitNetExplorer. By gaining more experience, it will become clear for what types of applications CitNetExplorer is most useful. We also expect to learn more about the strengths and weaknesses of different ways in which citation networks can be analyzed. For instance, in order to delineate the literature on a research topic, one may take either a clustering approach (like we do in our analysis of the science mapping literature in Subsection 3.2) or an approach involving a combination of keyword search and citation-based expansion (like we do in our analysis of the community detection literature in Subsection 3.3). CitNetExplorer can be used to test both approaches and to compare their results.

Based on our own experience with CitNetExplorer and the feedback that we receive from others, we plan to continue the development of the tool. Among other things, we will consider the possibility of including additional options for analyzing citation networks, for instance related to the idea of main path analysis (Hummon & Doreian, 1989). We especially hope that CitNetExplorer will contribute to a better understanding of the way in which research fields develop over time. We very much welcome suggestions for improvements and extensions of CitNetExplorer.

## Acknowledgment

We thank Katy Börner, Birger Larsen, Loet Leydesdorff, Erjia Yan, and our colleagues at the Centre for Science and Technology Studies, in particular Clara Calero, Rodrigo Costas, Jos Winnink, and Alfredo Yegros, for their input and feedback during the development of CitNetExplorer. We thank Ton van Raan for his help in interpreting the physics citation network.## References